\documentclass[preprint,notoc]{JHEP3}
\usepackage{graphics,slashed,algorithm,algorithmic,url}

\newcommand{\nn}{\nonumber \\}
\newcommand{\el}{\nonumber \\ &&}

\newcommand{\M}{\mathcal{M}}

\newenvironment{changemargin}[2]{%
  \begin{list}{}{%
    \setlength{\topsep}{0pt}%
    \setlength{\leftmargin}{#1}%
    \setlength{\rightmargin}{#2}%
    \setlength{\listparindent}{\parindent}%
    \setlength{\itemindent}{\parindent}%
    \setlength{\parsep}{\parskip}%
  }%
  \item[]}{\end{list}}
\floatstyle{plain} 
\newfloat{myalgo}{tbhp}{mya}  
\newenvironment{Algorithm}[1][tbh]
{\begin{myalgo}[#1] \centering \begin{minipage}{16cm} \begin{algorithm}[H]}
{\end{algorithm} \end{minipage} \end{myalgo}}

\newcommand{\be}{\begin{equation}} 
\newcommand{\ee}{\end{equation}} 
\newcommand{\cmb}{\begin{changemargin}}
\newcommand{\cme}{\end{changemargin}}
\newcommand{\bea}{\begin{eqnarray}} 
\newcommand{\eea}{\end{eqnarray}}

\def\spa#1.#2{\langle#1\,#2\rangle}
\def\spb#1.#2{[#1\,#2]}
\def\sandmm#1.#2.#3{%
\left\langle\smash{#1}{\vphantom1}\right|{#2}%
\left|\smash{#3}{\vphantom1}\right]}
\def\spab#1.#2.#3{\sandmm#1.#2.#3}
\def\spba#1.#2.#3{\sandpp#1.#2.#3}
\def\spaa#1.#2.#3.#4{\sandmp#1.{#2#3}.#4}
\def\spbb#1.#2.#3.#4{\sandpm#1.{#2#3}.#4}
\def\spash#1.#2{\spa{\smash{#1}}.{\smash{#2}}}
\def\spbsh#1.#2{\spb{\smash{#1}}.{\smash{#2}}}

\def\ksl{\not{\hbox{\kern-2.3pt $k$}}}

\def\e{\epsilon}

\preprint{IFT-UAM/CSIC-11-89}

\title{A New Algorithm For The Generation Of Unitarity-Compatible Integration By Parts Relations}

\author{Robert M. Schabinger\\
Instituto de F\'{i}sica Te\'{o}rica UAM/CSIC and \\
Departamento de F\'{i}sica Te\'{o}rica,\\
	Universidad Aut\'{o}noma de Madrid,\\
        Cantoblanco, E-28049 Madrid, Espa\~{n}a}
        
\abstract{Many multi-loop calculations make use of integration by parts relations to reduce the large number of complicated Feynman integrals that arise in such calculations to a simpler basis of master integrals. Recently, Gluza, Kajda, and Kosower argued that the reduction to master integrals is complicated by the presence of integrals with doubled propagator denominators in the integration by parts relations and they introduced a novel reduction procedure which eliminates all such integrals from the start. Their approach has the advantage that it automatically produces integral bases which mesh well with generalized unitarity. The heart of their procedure is an algorithm which utilizes the weighty machinery of computational commutative algebra to produce complete sets of unitarity-compatible integration by parts relations. In this paper, we propose a conceptually simpler algorithm for the generation of complete sets of unitarity-compatible integration by parts relations based on recent results in the mathematical literature. A striking feature of our algorithm is that it can be described entirely in terms of straightforward linear algebra.}


\dedicated{\centerline{Submitted to JHEP}}

\begin{document}

%%%%%%%%%%%%%%%%%%%%%%%%%%%%%%%%%%%%%%%%%%%%%%%%%%%
\tableofcontents
\section{Introduction}
\label{IntroSection}
When the technique of integration by parts in $d$ dimensions was first proposed by Tkachov and Chetyrkin~\cite{Chetyrkin}, it represented a major breakthrough in the study of perturbative gauge theories at the multi-loop level.\footnote{For the reader less familiar with integration by parts, we strongly recommend Smirnov's excellent book on Feynman integral calculus~\cite{EFInts}. He explains the technique and works through a number of simple (and non-simple) examples.} Their discovery was of fundamental practical importance, as it allowed researchers to perform many multi-loop calculations that were previously thought to be intractable. Furthermore, the idea of the technique is simple to describe. By taking integrals of derivatives in $d$ dimensions, one generates a tower of equations for the Feynman integrals belonging to a particular topology. Then one tries to solve these equations, either by inspection or via some systematic procedure, for an independent basis of master integrals. 

Unfortunately, the total number of equations generated in this way grows rapidly with the number of loops and external states in the integral topology under consideration. As a consequence, the solution of the so-called integration by parts relations is complicated for all but the simplest examples. For many years after the technique was introduced, no systematic procedure for the solution of integration by parts relations was known and it was therefore only possible to apply the method to simple integral topologies.\footnote{Of course, simple is a relative term. For a rather impressive early application of the method at the three-loop level see the long write-up for {\sc MINCER}~\cite{MINCER}.} The situation changed just over a decade ago with the introduction of a Gaussian elimination-like solution algorithm. This algorithm, due to Laporta~\cite{Laporta}, was a crucial step forward because it allowed researchers to apply the integration by parts technique to highly non-trivial problems for which an {\it ad hoc} approach would be impractical if not impossible. Although Laporta's algorithm has been tremendously successful, it has long been known that it may lead to master integrals with doubled propagator denominators.

To be more precise, for an $L$-loop topology, let $V = \{\ell_1,\dots,\ell_L,k_1\dots,k_{N}\}$ be the set of loop momenta together with a set of $N$ independent external momenta and let $v$ be a generic element of this set. Normally, one generates the set of integration by parts relations in an obvious way, considering each $v \in V$ in turn and writing all possible equations of the form
\bea
0 = \prod_{i = 1}^L\left( \int {d^d \ell_i\over (2 \pi)^d}\right)~ {\partial\over \partial \ell_j}\cdot\left( {v^{(j)}~ \mathcal{N}_1(\ell_1,\dots\ell_L)^{a_1}\cdots \mathcal{N}_q(\ell_1,\dots\ell_L)^{a_q}\over D_1(\ell_1,\dots\ell_L)^{b_{1}}\cdots D_m(\ell_1,\dots\ell_L)^{b_{m}}} \right)\,.
\label{IBP1}
\eea
Crucially, the indices $a_i$ and $b_i$ satisfy certain boundary conditions; typically the irreducible numerators (here we assume that each $\mathcal{N}_i$ has mass dimension two) for a given topology enter raised to, at most, some relatively small positive integer and, in some cases, certain propagator denominators are constrained to have non-negative indices since otherwise the resulting integrals vanish in dimensional regularization. This set of equations together with the appropriate boundary conditions can then be fed into Laporta's algorithm. The price one pays for being able to trivially generate the set of integration by parts relations in this fashion is twofold. Not only will the algorithm (Laporta or typical variation thereof~\cite{AIR,FIRE,REDUZE}) used to solve the resulting system of equations typically have to eliminate an enormous number of spurious integrals as it attempts to solve the system, it is not straightforward to ensure that each integral in the basis ultimately returned by the algorithm has the property that $b_i = 1$. 
 
Although there is nothing wrong with integrals that have some $b_i > 1$, they may be inconvenient for particular multi-loop applications. For example, in computational approaches based on generalized unitarity, one would like to have a basis with well-defined unitarity cuts in all channels (see {\it e.g.}~\cite{KLunitarity} and~\cite{MOunitarity} for recent work in this direction at two loops). It is unclear how one would make sense of integrals with doubled propagator denominators in such a framework. Gluza, Kajda, and Kosower (hereafter referred to as GKK) also argued in~\cite{GKK} that master integrals with doubled propagator denominators can be significantly harder to expand in $\e$ than those without.

The idea of the GKK procedure is relatively easy to state now that the stage is set. With the above motivation, GKK found that they could completely avoid the introduction of doubled propagators by imposing $b_i = 1$ from the beginning. They observed that, generically, one expects doubled propagators for the simple reason that the derivatives in eqs. (\ref{IBP1}) act on the propagator denominators, $D_k$. They also recognized that there is no good reason why one ought to consider the elements of $V$ one at a time; one can generalize eqs. (\ref{IBP1}) by replacing $v^{(j)}$ with a general linear combination of the elements of $V$:
\bea
0 = \prod_{i = 1}^L\left( \int {d^d \ell_i\over (2 \pi)^d}\right)~ {\partial\over \partial \ell_j}\cdot\left( {\sum_{i = 1}^{L + N}\alpha_i^{(j)} v_i^{(j)}~ \mathcal{N}_1(\ell_1,\dots\ell_L)^{a_1}\cdots \mathcal{N}_q(\ell_1,\dots\ell_L)^{a_q}\over D_1(\ell_1,\dots\ell_L)^{b_{1}}\cdots D_m(\ell_1,\dots\ell_L)^{b_{m}}} \right)\,.
\eea
It is also convenient at this point to combine together some of the equations by summing over $j$:
\cmb{-.2 in}{0 in}
\bea
0 = \prod_{i = 1}^L\left( \int {d^d \ell_i\over (2 \pi)^d}\right)~\sum_{j=1}^L {\partial\over \partial \ell_j}\cdot\left( {\sum_{i = 1}^{L + N}\alpha_i^{(j)} v_i^{(j)}~ \mathcal{N}_1(\ell_1,\dots\ell_L)^{a_1}\cdots \mathcal{N}_q(\ell_1,\dots\ell_L)^{a_q}\over D_1(\ell_1,\dots\ell_L)^{b_{1}}\cdots D_m(\ell_1,\dots\ell_L)^{b_{m}}} \right)\,.
\eea
\cme

In a nutshell, the GKK strategy is to start with all the $b_i$ equal to one
\bea
0 = \prod_{i = 1}^L\left( \int {d^d \ell_i\over (2 \pi)^d}\right)~\sum_{j=1}^L {\partial\over \partial \ell_j}\cdot\left( {\sum_{i = 1}^{L + N}\alpha_i^{(j)} v_i^{(j)}~ \mathcal{N}_1(\ell_1,\dots\ell_L)^{a_1}\cdots \mathcal{N}_q(\ell_1,\dots\ell_L)^{a_q}\over D_1(\ell_1,\dots\ell_L)\cdots D_m(\ell_1,\dots\ell_L)} \right)\,
\label{IBP2}
\eea
and then choose the coefficients $\alpha_i^{(j)}$ in such a way that the numerator exactly cancels all unwanted, derivative-generated powers of the propagator denominators. In other words, for each $k$, we demand that
\be
 \sum_{j = 1}^L \sum_{i = 1}^{L + N} \alpha_i^{(j)}\,{\partial D_k \over \partial \ell_j}\cdot v_i^{(j)}~ \mathcal{N}_1(\ell_1,\dots\ell_L)^{a_1}\cdots \mathcal{N}_q(\ell_1,\dots\ell_L)^{a_q} = - 2 T^{(k,a_1,\dots,a_q)} D_k \,,
\label{IBP3}
\ee
where $T^{(k,a_1,\dots,a_q)}$ is some polynomial built out of the irreducible numerators and Lorentz invariant combinations of the vectors in $V$.\footnote{We write $- 2 T^{(k,a_1,\dots,a_q)}$ instead of $T^{(k,a_1,\dots,a_q)}$ so that we will ultimately arrive at the same form that GKK did in their eqs. (4.1).} Each independent set of $L (L + N)$ coefficients $\alpha_i^{(j)}$ satisfying eqs. (\ref{IBP3}), upon substitution into eqs. (\ref{IBP2}), yields a unitarity-compatible integration by parts relation, by construction free of doubled propagators.

The downside of this novel approach is that now one needs to find a way to generate complete sets of $\alpha_i^{(j)}$  coefficients and it turns out that this is not a straightforward task. Although GKK did propose a solution\footnote{Actually, GKK presented two distinct algorithms for the generation of complete sets of unitarity-compatible integration by parts relations. In what follows, when we refer to GKK's ``solution'' or ``solution algorithm'' with no additional qualifier, we are referring to their best solution (what they call Algorithm III).} to this problem in~\cite{GKK}, they admitted that their solution was somewhat provisional and that there is likely room for improvement. The GKK solution relies heavily on the use of  Gr\"{o}bner bases, important constructs in computational commutative algebra which are, however, notoriously difficult to compute in practice~\cite{alggeom}.\footnote{For the reader less well-versed in computational commutative algebra, we can strongly recommend the very well-written and concise survey by Adams and Loustaunau~\cite{AL}. Most of the relevant mathematical concepts are also defined and briefly explained in the GKK paper~\cite{GKK}.} In this paper we propose a completely different solution to the problem of generating complete sets of unitarity-compatible integration by parts relations. As we shall see, our Algorithm \ref{alg} is  based entirely on simple linear algebra and, in particular, completely avoids the use of Gr\"{o}bner bases.

This article is organized as follows. In section~\ref{review} we describe more precisely the problem we wish to solve and introduce some useful notation. In section~\ref{body} we present the main result of this paper, Algorithm \ref{alg}, and talk the reader through it. In Section \ref{example}, we give a detailed example of how our algorithm works in practice. In section~\ref{ConclusionSection} we present our conclusions and outline our plan for future research.
%%%%%%%%%%%%%%%%%%%%%%%%%%%%%

\section{Preliminaries}
\label{review}
In this section we take a closer look at the reduction procedure introduced by GKK and discuss its strengths and weaknesses. Our initial goal will be to precisely set up the mathematical problem that lies at the heart of the GKK procedure and discuss why (in the opinion of both GKK and the present author) the solution presented in~\cite{GKK} is not likely to be the best one possible. We will then explain our approach to the problem and illustrate with a very simple example what precisely Algorithm \ref{alg} is designed to do.

As we left them, eqs. (\ref{IBP3}) look rather cumbersome. We can clean them up by absorbing the numerator polynomial $\mathcal{N}_1(\ell_1,\dots\ell_L)^{a_1}\cdots \mathcal{N}_q(\ell_1,\dots\ell_L)^{a_q}$ into the $\alpha_i^{(j)}$
\be
\sum_{j = 1}^L\sum_{i = 1}^{L + N} \alpha_i^{(j)}\,{1 \over 2}{\partial D_k \over \partial \ell_j}\cdot v_i^{(j)} + T^{(k, \alpha)} D_k = 0 \,,
\label{IBP4}
\ee
with the understanding that now the $\alpha_i^{(j)}$ are dimensionful. We will have to take this into account in our search for independent solutions. To make further progress, GKK expressed (\ref{IBP4}) as a matrix equation:
\be
\beta \cdot E = {\bf 0}\,,
\label{IBP5}
\ee
where
\be
\beta = \Big(\alpha_1^{(1)},\dots,\alpha_{L+N}^{(1)},\dots\dots,\alpha_1^{(L)},\dots,\alpha_{L+N}^{(L)},T^{(1, \alpha)},\dots,T^{(m, \alpha)}\Big)\,.
\ee
Given an explicit expression for $V$ and an ordering on the set of propagator denominators, the entries of $E$ can be straightforwardly read off from (\ref{IBP4}). For an explicit example, we refer the reader to eq. (5.3) of~\cite{GKK}. In eq. (5.3), GKK wrote out $E$ explicitly for the planar massless double box, choosing $\{\ell_1^2, \ell_1^2 - 2\ell_1\cdot k_1, \ell_1^2 - 2\ell_1\cdot k_1 -2 \ell_1\cdot k_2 +s_{1 2}, \ell_2^2, \ell_2^2 - 2 \ell_2\cdot k_4, \ell_2^2 - 2 \ell_2\cdot k_3 - 2 \ell_2\cdot k_4 + s_{12},\ell_1^2 + \ell_2^2 + 2 \ell_1\cdot \ell_2\}$ for the ordering on the set of propagator denominators (this ordering fixes the sequence of the columns of $E$).\footnote{We should also point out that the precise definitions GKK made for the $\alpha_i^{(j)}$ (what they call $c_j^{(i)}$) and $T^{(k,\alpha)}$ (what they call $u_k$) do not seem to exactly reproduce eq. (5.3). However, the differences can be taken into account by appropriately rescaling the unknowns and are therefore unimportant.} The point of making this rearrangement is that now the problem of determining all independent sets of $\alpha_i^{(j)}$ coefficients looks like a well-known, well-studied problem from computational commutative algebra: the computation of the syzygy module\footnote{Given a set of generators for a module $\mathcal{M}$, $\{{\bf M}_1,\dots,{\bf M}_r\}$, the syzygy module of $\mathcal{M}$ is simply the set of all $\beta = (b_1,\dots,b_r)$ such that $\sum_{i = 1}^r b_i \cdot {\bf M}_i = {\bf 0}$. In this paper, we will deal only with submodules of $F[x_1,\dots,x_n]^r$ and $F[x_1,\dots,x_n]^m$ for a field, $F$.} of a module for which one has an explicit set of generators.

Actually, as explained by GKK, it suffices to solve this problem for ideals, since given a set of generators for a module $\mathcal{M}$, one can easily define a set of generators for an ideal with exactly the same syzygies. Suppose the set $\{{\bf M}_1,\dots,{\bf M}_r\}$ generates $\mathcal{M}$. By simply taking the dot product of each ${\bf M}_i$ with a tuple of dummy variables satisfying the relations $t_i t_j = 0$, $\{t_1,\dots,t_m\}$, we can convert our generating set for $\M$ into a generating set for an ideal $\mathcal{I}$ canonically associated to $\M$. The class of ideals canonically associated to the modules described by eqs. (\ref{IBP4}) can be taken to be homogeneous of uniform degree two. We can see this as follows. By definition, each parameter entering into the matrix $E$ of eq. (\ref{IBP5}) is of the form ${\partial D_k \over \partial \ell_j}\cdot v_i^{(j)}$ or $D_k$ and is going to have mass dimension two. Therefore, it makes sense to take every independent Lorentz product that can arise in $E$ to be an independent variable and the class of modules described by eqs. (\ref{IBP4}) to be homogeneous of uniform degree one. It then follows, after applying the canonical map described above, that the ideals of interest are homogeneous of uniform degree two. In the end, we find that each generator of $\mathcal{I}$ has terms of the schematic form $a\, t_j\,x_k$, $a \in F$ for some field $F$.\footnote{In this paper, $F$ is given by the field of rational functions of the auxiliary parameters in the matrix $E$ ({\it e.g.} $\chi_{14}$ in the case of the planar massless double box treated in detail by GKK) of eq. (\ref{IBP5}). However, for practical purposes, it is more convenient to simply assign prime numbers to the dimensionless auxiliary parameters and work over $\mathbb{Q}$.} 

Before continuing, we need to introduce a little more notation and point out an important fact about the syzygies of generating sets for homogeneous ideals of uniform degree. Given an ideal generated by $P = (p_1,\dots,p_r)$, we define $S_d(P)$ to be the set of all syzygies of $P$ such that $\beta =  (b_1,\dots,b_r) \in S_d(P)$ implies that each $b_i$ is a polynomial of uniform degree $d$. We will typically refer to $S_d(P)$ as the set of all degree $d$ syzygies of $P$. It turns out that, for the case of homogeneous ideals of uniform degree, the syzygy module, $S(P)$, is a graded module~\cite{LASyz}:
\be
S(P) = \bigoplus_{d = 0}^\infty S_d(P)\,.
\ee
This means that, for homogeneous ideals of uniform degree, it suffices to search for degree $d$ syzygies.

Determining a basis for the syzygy module of a generic ideal is known to be a very difficult problem~\cite{alggeom} (and is very much an active area of mathematical research). Therefore, one needs a dedicated solution algorithm, tailor-made for the class of ideals described above. As mentioned in the introduction, the solution algorithm employed by GKK relies heavily on the use of Gr\"{o}bner bases. GKK chose the well-known Buchberger algorithm~\cite{Buchberger} to compute their Gr\"{o}bner bases. In their paper, GKK pointed out that there have been a number of recent attempts to improve on Buchberger's algorithm (see {\it e.g.}~\cite{Eder} for a description of one of the most promising of these recent attempts, based on Faug\`{e}re's $F_5$ algorithm~\cite{Faugere}) and concluded that their Buchberger-based approach was not likely to be optimal. In this paper we rethink their approach at a more fundamental level. 

Certainly, one could attempt to compute complete sets of syzygies using an approach based on Faug\`{e}re's $F_5$ algorithm~\cite{SyzF5} or some other improved algorithm for the computation of Gr\"{o}bner bases. However, it is actually no longer clear that one should use Gr\"{o}bner bases at all. Quite recently (after the appearance of~\cite{GKK}) it was shown in~\cite{LASyz} that, remarkably, bases for the modules of syzygies of special classes of ideals can be computed using simple linear algebra. Actually, at this stage, the reader may be wondering what stopped GKK from using a linear algebra-based approach in the first place. Na\"{i}vely, linear algebra seems to offer a very easy way to compute syzygies. 

To illustrate why the ideas of reference~\cite{LASyz} are non-trivial, we consider the ideal generated by $P = \{x_1 - x_2, 2 x_2 - x_1\}$ and attempt to compute its syzygies using linear algebra.\footnote{For the sake of definiteness, suppose that we are working with polynomials in $\mathbb{Q}[x_1,x_2]$.} By inspection, we see that $P$ has no degree zero syzygies. By definition, a degree one syzygy of $P$ must have the form $(c_1 x_1 + c_2 x_2, c_3 x_1 + c_4 x_2)$ for some elements $(c_1,c_2,c_3,c_4)$ of $\mathbb{Q}$. Starting with this ansatz, we can take the dot product with $P$
\be
(c_1 x_1 + c_2 x_2, c_3 x_1 + c_4 x_2)\cdot P = (c_1 - c_3) x_1^2 + (-c_1 + c_2 + 2 c_3 - c_4) x_1 x_2 + (-c_2 + 2 c_4) x_2^2\,,
\ee
set the coefficient of each monomial in the sum to zero, and solve the resulting system of equations. It turns out that there is a one-parameter family of solutions which we parametrize by $c_4$:
\be
(c_1,c_2,c_3,c_4) = (-c_4, 2 c_4, -c_4, c_4)\,.
\ee
We can arbitrarily set $c_4 = 1$ to obtain a basis for $S_1(P)$, $(- x_1 + 2 x_2, - x_1 + x_2)$. So far so good. By definition, a degree two syzygy of $P$ must have the form $(c_1 x_1^2 + c_2 x_1 x_2 + c_3 x_2^2, c_4 x_1^2 + c_5 x_1 x_2 + c_6 x_2^2)$. Going through the same procedure, we arrive at a two-parameter family of solutions which we parametrize by $c_5$ and $c_6$:
\be
(c_1,c_2,c_3,c_4,c_5,c_6) = (- c_5 - c_6, 2 c_5 + c_6, 2 c_6, - c_5 - c_6, c_5, c_6)\,.
\ee
By replacing $(c_5,c_6)$ with each of the standard basis vectors for $\mathbb{R}^2$ in turn, we find that $(-1, 2, 0, -1, 1, 0)$ and $(-1, 1, 2, -1, 0, 1)$ form a basis for the set of solutions. In order to map these solutions back to degree two syzygies of $P$, we partition each vector into two non-overlapping subsets of length three without changing the overall ordering of the entries 
\be
\Big\{\Big(\{-1,2,0\},\{-1,1,0\}\Big), \Big(\{-1,1,2\},\{-1,0,1\}\Big)\Big\}
\ee
and then dot each resulting 3-tuple into $\{x_1^2, x_1 x_2, x_2^2\}$:
\bea
&&\Big\{\Big(\{x_1^2, x_1 x_2, x_2^2\}\cdot\{-1,2,0\},\{x_1^2, x_1 x_2, x_2^2\}\cdot\{-1,1,0\}\Big),
\el\Big(\{x_1^2, x_1 x_2, x_2^2\}\cdot\{-1,1,2\},\{x_1^2, x_1 x_2, x_2^2\}\cdot\{-1,0,1\}\Big)\Big\} 
\el= \Big\{\Big(- x_1^2 + 2 x_1 x_2, - x_1^2 + x_1 x_2\Big), \Big(- x_1^2 + x_1 x_2 + 2 x_2^2, - x_1^2 + x_2^2\Big)\Big\}\,.
\label{examp1}
\eea
This overly formal description of the mapping back to syzygies of $P$ is overkill for the example at hand but will be useful later on. 

We might be tempted to conclude that we have found two new, linearly independent, degree two syzygies of $P$. This, however, is not true. It turns out that the syzygies of eq. (\ref{examp1}) can be expressed as multiples of  $(- x_1 + 2 x_2, - x_1 + x_2)$. Explicitly, we have
\be
\Big(- x_1^2 + 2 x_1 x_2, - x_1^2 + x_1 x_2\Big) = x_1 (- x_1 + 2 x_2, - x_1 + x_2)
\ee
and
\be
\Big(- x_1^2 + x_1 x_2 + 2 x_2^2, - x_1^2 + x_2^2\Big) = (x_1 + x_2)(- x_1 + 2 x_2, - x_1 + x_2)\,.
\ee

Besides highlighting the profound difference between linear independence in a vector space and linear independence in a module, this example shows what goes wrong if one tries to compute the syzygies of homogeneous ideals of uniform degree using na\"{i}ve linear algebra. One is able to compute syzygies in a straightforward manner but not a basis of linearly independent syzygies. It is therefore remarkable that, under appropriate assumptions, this obstacle is actually surmountable. In the next section, drawing upon some of the ideas introduced in~\cite{LASyz}, we present a simpler linear algebra-based alternative to the solution adopted by GKK.
\section{The Algorithm}
\label{body}
The purpose of this section is to give an explicit pseudo-code detailing our solution to the problem defined in Section \ref{review} and to thoroughly comment it. The pseudo-code we present (Algorithm \ref{alg} and associated subroutines) is quite explicit and should allow the reader to fashion a crude implementation of our algorithm in {\tt Maple} or {\tt Mathematica} with very little effort (beyond that required to understand the algorithm in the first place). We should emphasize that we do not in any way claim that an implementation based on our pseudo-code is an {\it effective} implementation. The pseudo-code given below is intended to be maximally clear as opposed to maximally efficient.\footnote{For example, the first statement in the upper branch of the If statement in Algorithm \ref{alg} is completely superfluous and was included only because, in our opinion, it makes the functionality of Subroutine \ref{sub1} significantly easier to understand.} Before discussing the non-trivial features of Algorithm \ref{alg}, we need to make a few more definitions. 

First, let $M_d^{(n)} = \Big\{X_1^{(d)},\dots,X_{\frac{(d+n-1)!}{d!(n-1)!}}^{(d)}\Big\}$ be the set of all monomials of degree $d$ built out of the $n$ variables $\{x_1,\dots,x_n\}$.\footnote{Note that there are precisely $\frac{(d+n-1)!}{d!(n-1)!}$ monomials of degree $d$ built out of $n$ variables. This result follows immediately once one recognizes that the monomial counting problem is isomorphic to one of the usual ``ball-box'' counting problems of enumerative combinatorics. We refer the unfamiliar reader to Section 1.4 of Stanley's textbook on the subject~\cite{Stanley}.} For example, $M_2^{(2)} = \{x_1^2, x_1 x_2, x_2^2\}$. For definiteness, we will always order sets of monomials lexicographically. This choice of ordering is just a choice and has no deeper significance. Now, if we have in hand a sequence of $r$ homogeneous polynomials of uniform degree two, $P_0 = \{p_1,\dots,p_r\}$, then $P_d$ is defined to be the outer product of $M_d^{(n)}$ and $P_0$:
\be
P_{d} = \Big\{p_1 X_1^{(d)},\dots,p_1 X_{\frac{(d+n-1)!}{d!(n-1)!}}^{(d)},\dots\dots,p_r X_1^{(d)},\dots,p_r X_{\frac{(d+n-1)!}{d!(n-1)!}}^{(d)}\Big\}\,.
\ee 
Clearly, $P_d$ is a set of $\frac{r (d+n-1)!}{d!(n-1)!}$ homogeneous polynomials of uniform degree $d + 2$. For example, if we take $P_0 = P = \{x_1 - x_2, 2 x_2 - x_1\}$ it is easy to see that $P_2 = \{(x_1 - x_2)x_1^2, (x_1 - x_2)x_1 x_2, (x_1 - x_2)x_2^2, (2 x_2 - x_1)x_1^2, (2 x_2 - x_1)x_1 x_2, (2 x_2 - x_1)x_2^2\}$. The construction of $P_2$ can be thought of as an intermediate step towards the extraction of the degree two syzygies of $P$. Instead of solving the system $(c_1 x_1^2 + c_2 x_1 x_2 + c_3 x_2^2, c_4 x_1^2 + c_5 x_1 x_2 + c_6 x_2^2)\cdot \{x_1 - x_2, 2 x_2 - x_1\} = 0$, we can construct $P_2$ and then solve the system  $(c_1,c_2,c_3,c_4,c_5,c_6)\cdot \{(x_1 - x_2)x_1^2, (x_1 - x_2)x_1 x_2, (x_1 - x_2)x_2^2, (2 x_2 - x_1)x_1^2, (2 x_2 - x_1)x_1 x_2, (2 x_2 - x_1)x_2^2\} = 0$. So, instead of trying to work with $S_d(P_0)$ directly, we can work with the vector space $S_0(P_d)$. This natural correspondence between basis vectors of $S_0(P_d)$ and degree $d$ syzygies of $P_0$ is an essential part of Algorithm \ref{alg}, which is why we took the time to carefully describe it while working through the illustrative example at the end of Section \ref{review}. However, it is important to remember that this map does not necessarily yield independent elements of $S(P_0)$; to work as advertised, our solution algorithm must be able to determine, using only linear algebra, what basis vectors of $S_0(P_d)$ correspond to new syzygies of $P_0$, linearly independent of those already determined.
\newpage
\begin{Algorithm}[H]
\caption{}\label{alg}
\begin{algorithmic}
\REQUIRE A minimal generating set for a homogeneous ideal of uniform degree two, $P_0 = \{p_1(t_1,\dots,t_m,x_1,\dots,x_n),\dots,p_r(t_1,\dots,t_m,x_1,\dots,x_n)\}$. The $p_i$ are further constrained to have terms of the schematic form $a\, t_j\,x_k$ with $a \in F$ and $t_i t_j = 0$ for all $i$ and $j$. ${\rm SyzList}_0 = B_0 = \{{\bf 0}\}$. For a given integral topology with generic irreducible numerator $\mathcal{N}_1(\ell_1,\dots\ell_L)^{a_1}\cdots \mathcal{N}_q(\ell_1,\dots\ell_L)^{a_q}$, $\Delta$ is the largest value of $\sum_{i = 1}^q a_i$ allowed by the boundary conditions on the $a_i$ (we assume that each $\mathcal{N}_i$ has mass dimension two). $M_d^{(n)} = \Big\{X_1^{(d)},\dots,X_{\frac{(d+n-1)!}{d!(n-1)!}}^{(d)}\Big\}$ is the lexicographically ordered set of all monomials of degree $d$ built out of the $n$ variables $\{x_1,\dots,x_n\}$.
\STATE
\STATE $\bullet$Set $d = 1$
\WHILE{$d \leq \Delta$}
\IF{$B_{d-1} \neq \{{\bf 0}\}$}
\STATE $\bullet$Take the outer product of $P_0$ and $M_{d-1}^{(n)}$ to determine $P_{d-1}$: \\~~~\,$P_{d-1} = \Big\{p_1 X_1^{(d-1)},\dots,p_1 X_{\frac{(d+n-2)!}{(d-1)!(n-1)!}}^{(d-1)},\dots\dots,p_r X_1^{(d-1)},\dots,p_r X_{\frac{(d+n-2)!}{(d-1)!(n-1)!}}^{(d-1)}\Big\}$
\STATE $\bullet$\,Take the outer product of $P_0$ and $M_{d}^{(n)}$ to determine $P_{d}$: \\~~~\,$P_{d} = \Big\{p_1 X_1^{(d)},\dots,p_1 X_{\frac{(d+n-1)!}{d!(n-1)!}}^{(d)},\dots\dots,p_r X_1^{(d)},\dots,p_r X_{\frac{(d+n-1)!}{d!(n-1)!}}^{(d)}\Big\}$
\STATE $\bullet$Apply Subroutine \ref{sub1} to each element of $B_{d-1}$ and call the union of the results $A$:
\\~~~\,$A = \bigcup\limits_{i=1}^{|B_{d-1}|} \sigma(\alpha_i)$
\STATE $\bullet$Regard the elements of $A$ as the rows of a matrix, find a row echelon form of this 
\\~\, matrix, keep only rows with at least one non-zero entry, and call the result $C$
\STATE $\bullet$Replace with zero the $s$ entries of $P_d$ which correspond
\\~\,to the pivot columns of $C$ and call the result $G$ 
\STATE $\bullet$Apply Subroutine \ref{sub2} to $G$ and call the result $D$: 
\\~~~\,$D = {\rm SSyz}(G)$
\STATE $\bullet$If $D \neq \{{\bf 0}\}$ then $B_d = C ~\bigcup~ D$ and if $D = \{{\bf 0 }\}$ then $B_d = C$
\STATE $\bullet$Apply Subroutine \ref{sub3} to $D$ to determine ${\rm SyzList}_d$:
\\~~~\,${\rm SyzList}_d = {\rm ToSyz}(D)$
\ELSE
\STATE $\bullet$\,Take the outer product of $P_0$ and $M_{d}^{(n)}$ to determine $P_{d}$: \\~~~\,$P_{d} = \Big\{p_1 X_1^{(d)},\dots,p_1 X_{\frac{(d+n-1)!}{d!(n-1)!}}^{(d)},\dots\dots,p_r X_1^{(d)},\dots,p_r X_{\frac{(d+n-1)!}{d!(n-1)!}}^{(d)}\Big\}$
\STATE $\bullet$Apply Subroutine \ref{sub2} to $P_d$ and call the result $D$: 
\\~~~\,$D = {\rm SSyz}(P_d)$
\STATE $\bullet B_d = D$
\STATE $\bullet$Apply Subroutine \ref{sub3} to $D$ to determine ${\rm SyzList}_d$:
\\~~~\,${\rm SyzList}_d = {\rm ToSyz}(D)$
\ENDIF
\STATE $\bullet d = d+1$
\ENDWHILE
\RETURN $\left(\bigcup\limits_{d=1}^\Delta~ {\rm SyzList}_d \right)\Big\backslash \{ {\bf 0}\}$
\end{algorithmic}
\end{Algorithm}
\floatname{algorithm}{Subroutine}
\addtocounter{algorithm}{-1}
\begin{Algorithm}[h!]
\caption{${\rm \sigma}(\alpha)$ maps $\alpha \in B_{d-1}$ to $n$ linearly independent elements of $S_0(P_d)$}\label{sub1}
\begin{algorithmic}
\REQUIRE The objects introduced in Algorithm \ref{alg} and $\alpha = \Big(A_1,\dots,A_{r(d+n-2)!\over (d-1)!(n-1)!}\Big)$
\\\qquad\qquad $= A_1 p_1 X_1^{(d-1)} +  \cdots + A_{\frac{r(d+n-2)!}{(d-1)!(n-1)!}} p_r X_{\frac{(d+n-2)!}{(d-1)!(n-1)!}}^{(d-1)} \in B_{d-1}$
\STATE 
\STATE
\RETURN $\Big\{A_1 p_1 X_1^{(d-1)}x_1 +  \cdots + A_{\frac{r(d+n-2)!}{(d-1)!(n-1)!}} p_r X_{\frac{(d+n-2)!}{(d-1)!(n-1)!}}^{(d-1)} x_1,\dots\dots,A_1 p_1 X_1^{(d-1)} x_n + \cdots$
\\\qquad\qquad$\cdots + A_{\frac{r(d+n-2)!}{(d-1)!(n-1)!}} p_r X_{\frac{(d+n-2)!}{(d-1)!(n-1)!}}^{(d-1)} x_n\Big\}$
\end{algorithmic}
\end{Algorithm}
\begin{Algorithm}[h!]
\caption{${\rm SSyz}(G)$ computes a basis for $S_0(G)$}\label{sub2}
\begin{algorithmic}
\REQUIRE The objects introduced in Algorithm \ref{alg} and a sequence, $G = \{q_1,\dots,q_n\}$, of homogeneous polynomials in the variables $\{t_1,\dots,t_m,x_1,\dots,x_n\}$ with coefficients in $F$.
\STATE
\STATE $\bullet$ Introduce a sequence of $|G|$ dummy variables:
\\~~~\,$Y = \{y_1,\dots,y_{|G|}\}$
\STATE $\bullet$Expand the dot product $G\cdot Y$ and partition the terms in the sum into equivalence classes,
\\~\,declaring terms equivalent if and only if they contain the same monomial in the variables 
\\~\,$\{t_1,\dots,t_m,x_1,\dots,x_n\}$
\STATE $\bullet$Set the sum of terms in each equivalence class equal to zero, collect the resulting set of
\\~\,equations constraining the $y_i$, and drop any equations that are obviously redundant
\STATE $\bullet$Put the remaining equations into matrix form:
\\~~~\,$Q~Y = {\bf 0}$
\RETURN a basis for ${\rm Null}(Q)$
\end{algorithmic}
\end{Algorithm}
\begin{Algorithm}[h!]
\caption{${\rm ToSyz}(D)$ maps $\nu$ elements of $S_0(P_d)$ to $\nu$ elements of $S_d(P_0)$}\label{sub3}
\begin{algorithmic}
\REQUIRE The objects introduced in Algorithm \ref{alg} and a list, $D = \{d_1,\dots,d_\nu\}$, $d_i \in S_0(P_d)$.
\STATE
\STATE $\bullet$Partition each vector $d_i$ into $r$ distinct non-overlapping sequences of length ${(d+n-1)!\over d!(n-1)!}$, 
\\~\,taking care to preserve the order of the entries of each $d_i$:
\\~~~\,$d_i \rightarrow \left\{z_1^{(i)},\dots,z_r^{(i)}\right\}$
\RETURN $\Big\{\left(M_d^{(n)}\cdot z_1^{(1)},\dots,M_d^{(n)}\cdot z_r^{(1)}\right),\dots\dots,\left(M_d^{(n)}\cdot z_1^{(\nu)},\dots,M_d^{(n)}\cdot z_r^{(\nu)}\right)\Big\}$
\end{algorithmic}
\end{Algorithm}
We now describe how Algorithm \ref{alg} works in some detail. By assumption, the $m$ propagator denominators of the integral topology under consideration form a linearly independent set.\footnote{If this is not the case, one should first reduce to simpler integral topologies satisfying this condition before attempting to apply Algorithm \ref{alg}. GKK explain this procedure in some detail in~\cite{GKK}.} This implies that $P_0$ is a minimal generating set for the ideal under consideration and there are no non-trivial syzygies of degree zero. Since Algorithm \ref{alg} proceeds incrementally in the degree of the syzygies, it is convenient to define two bookkeeping lists, $B_d$ and ${\rm SyzList}_d$, indexed by $d$. $B_d$ is a basis for the vector space $S_0(P_d)$ and ${\rm SyzList}_d$ is a set of linearly independent elements of $S_d(P_0)$ which are, for $d > 1$, also linearly independent of all syzygies of degree $\leq d - 1$ in the set $\bigcup_{i = 1}^{d - 1} {\rm SyzList}_i$ (we will call the elements of ${\rm SyzList}_d$ new degree $d$ syzygies of $P_0$). Due to the fact that $P_0$ has no degree zero syzygies, both lists are initialized to $\{{\bf 0}\}$ and $d$ is initialized to one. Since $B_0 = \{{\bf 0}\}$, Algorithm \ref{alg} always starts with a pass through the lower branch of the If statement. Using Subroutine \ref{sub2}, Algorithm \ref{alg} determines a basis for $S_0(P_1)$ along the lines described in the example near the end of Section \ref{review}. For the sake of discussion, we assume that this basis is non-trivial. This basis, $D$, is then used to determine both $B_1$ and, via Subroutine \ref{sub3}, ${\rm SyzList}_1$. In this section we will not step into Subroutines \ref{sub2} or \ref{sub3}. There is nothing non-trivial about them and we will in any case go through the subroutines once explicitly in Section \ref{example}.

Increment $d$ to two. Subroutine \ref{sub1} lifts the syzygies in $B_1$ to syzygies in $S_0(P_2)$ in every independent way possible. This is accomplished by simply multiplying each of the elements of $B_1$ (written out explicitly in terms of their coordinates, the elements of $P_1$) by each one of the variables $\{x_1,\dots,x_n\}$ in turn. The results can then be interpreted as syzygies in $S_0(P_2)$ (written out explicitly in terms of their coordinates, the elements of $P_2$). Clearly, if $\beta$ is a scalar syzygy of $P_1$, $\beta\cdot (x_i\,P_1) = 0$ for each $1 \leq i \leq n$ by linearity. This implies that $\beta$ can be interpreted as a scalar syzygy of $P_2$ in $n$ independent ways. Given $\nu$ syzygies in $S_0(P_1)$, Subroutine \ref{sub1} produces $\nu\, n$ syzygies in $S_0(P_2)$ (some of which may be linearly dependent). As we shall see, Subroutine \ref{sub1} is a crucial first step towards determining which syzygies in $S_0(P_2)$ correspond to new degree two syzygies of $P_0$ and which do not.\footnote{The reader who has read reference~\cite{LASyz} might be concerned that we have not yet spoken about the principal syzygies of our ideals. Actually, the fact that the $t_i$ are simply coordinate variables with no independent existence of their own (they satisfy $t_i t_j = 0$ for all $i$ and $j$) implies that the ideals of interest to us have no principal syzygies at all.}

Since the syzygies in $S_0(P_2)$ produced by Subroutine \ref{sub1} are not guaranteed to be linearly independent, the next step is to put the output of Subroutine \ref{sub1} into row echelon form, discard all rows without non-zero entries, and call the result $C$. Since, by construction, each of the $s$ rows of $C$ is a scalar syzygy of $P_2$, we could in principle rewrite $s$ of the polynomials in $P_2$ as linear combinations of the other ${r(d + n - 1)!\over d! (n - 1)!} - s$. Algorithm \ref{alg} takes this fact into account by replacing the $s$ entries of $P_2$ which correspond to the pivot columns of $C$ with zero and calling the result $G$. In constructing $G$, it has isolated the subspace of $S_0(P_2)$ which is in correspondence with the new degree two syzygies of $P_0$ (if any new degree two syzygies exist). Next, Algorithm \ref{alg} uses Subroutine \ref{sub2} to determine a basis, $D$, for the subspace of $S_0(P_2)$ under consideration. However, in order to search for degree three syzygies in an analogous fashion, we need a basis for $S_0(P_2)$ in its entirety. Therefore, Algorithm \ref{alg} sets $B_2 = C \bigcup D$ (for the sake of discussion assuming that Subroutine \ref{sub2} found new degree two syzygies). Finally, Algorithm \ref{alg} solves for the new degree two syzygies themselves by applying Subroutine \ref{sub3} to $D$. After incrementing $d$ to three, Algorithm \ref{alg} would pass through the upper branch of the If statement again in an attempt to find new degree three syzygies of $P_0$. 

Of course, we have been assuming throughout our discussion that $\Delta > 3$. It is worth emphasizing that, the termination condition we are using for Algorithm \ref{alg} comes entirely from the physics. The fact that irreducible numerators typically have, at worst, a relatively small mass dimension is the only reason that we were able to fruitfully apply the ideas of reference~\cite{LASyz} and construct Algorithm \ref{alg}. Although we believe that the treatment given here is appropriate, some readers may prefer a more formal one. If that is the case, then we recommend reading~\cite{LASyz}. Most of the non-trivial aspects of our pseudo-code are treated there as well in the style preferred by mathematicians.
\section{An Explicit Example}
\label{example}
In this section, we show how Algorithm \ref{alg} works in practice by going through it for a particular example. It was challenging to find a physically motivated example compact enough to present in detail and, at the same time, rich enough to give the reader a good feeling for how the algorithm functions. In the end, we found that the module given by the irreducible part of the module associated (associated in the sense of the construction reviewed in Section \ref{review}) to the planar massless double box works very well. In the solution algorithm of GKK, the study of this module (hereafter referred to as $\mathcal{M}$) is the first step towards the determination of the complete set\footnote{Here it is perhaps worth pointing out that, in fact, the application of our algorithm to the module associated to the planar massless double box yields more independent solutions than GKK found (working modulo reducibility as they do). We conjecture that, perhaps, GKK did not really seek the {\it complete} set of linearly independent syzygies modulo reducibility but were instead content to determine a subset sufficient for the elimination of as many irreducible numerators as possible. If our reading of GKK is correct, then the obvious question is whether discarding potentially useful linear relations between Feynman integrals of a given topology is prudent. We suspect that this is not the best strategy because, at least for the planar double box, the elimination of irreducible numerators in this manner seems to lead to a large number of irreducible integrals of simpler topology. Unfortunately, further exploration of this interesting question is beyond the scope of the present paper.} of linearly independent syzygies of the module associated to the planar massless double box. In their paper, GKK assert that $\mathcal{M}$ has just three linearly independent syzygies: one of degree one and two of degree two. This {\it a priori} knowledge of the syzygy module of $\M$ will allow us to stop the example when it ceases to be interesting. Otherwise we would have to make several more trips through the While loop (as explained in~\cite{GKK}, $\Delta = 6$ for the planar massless double box), each time discovering no new syzygies. For the sake of clarity, our exposition will mirror the pseudo-code of Section \ref{body} quite closely.

As a preliminary step we must derive the generators of $\M$. In Section \ref{review}, we pointed out that the ordering adopted by GKK for the propagator denominators of the planar massless double box is given by $\{\ell_1^2, \ell_1^2 - 2\ell_1\cdot k_1, \ell_1^2 - 2\ell_1\cdot k_1 -2 \ell_1\cdot k_2 +s_{1 2}, \ell_2^2, \ell_2^2 - 2 \ell_2\cdot k_4, \ell_2^2 - 2 \ell_2\cdot k_3 - 2 \ell_2\cdot k_4 + s_{12},\ell_1^2 + \ell_2^2 + 2 \ell_1\cdot \ell_2\}$. We actually prefer the ordering $\{\ell_1^2, \ell_1^2 - 2\ell_1\cdot k_1, \ell_1^2 - 2\ell_1\cdot k_1 -2 \ell_1\cdot k_2 +s_{1 2}, \ell_2^2 - 2 \ell_2\cdot k_3 - 2 \ell_2\cdot k_4 + s_{12}, \ell_2^2 - 2 \ell_2\cdot k_4, \ell_2^2,\ell_1^2 + \ell_2^2 + 2 \ell_1\cdot \ell_2\}$ and this is what we will use. Note, however, that we do adopt their ordering for the generators themselves. By definition, the generators of $\M$ are the generators of the module associated to the planar massless double box by eqs. (\ref{IBP4}) (the rows of the matrix $E$ in eq. (5.3) of~\cite{GKK}) reduced over the propagator denominators of the massless double box. This reduction is effected by making the substitutions $\{k_3 \rightarrow - k_1 - k_2 - k_4, \ell_1^2 \rightarrow 0, \ell_1 \cdot k_1 \rightarrow 0, \ell_2^2 \rightarrow 0, \ell_2 \cdot k_4 \rightarrow 0, \ell_1 \cdot \ell_2 \rightarrow 0, \ell_1\cdot k_2 \rightarrow s_{12}/2,\ell_2\cdot k_2 \rightarrow -\ell_2\cdot k_1 - s_{12}/2\}$. We find that $\M$ is generated by 
\bea
&&\bigg\{\left(0,0,-\frac{s_{12}}{2},0,0,0,0\right), \left(0,-\ell_2\cdot k_1,\frac{s_{12}}{2},0,0,0,0\right), \left(0,0,-\frac{s_{12}}{2},0,0,0,\ell_2\cdot k_1\right),
\el
\left(\frac{s_{12}}{2},0,0,0,0,0,-\ell_2\cdot k_1\right), \left(\ell_1\cdot k_4,-\frac{\chi_{14}s_{12}}{2}+\ell_1\cdot k_4,\frac{s_{12}}{2}+\ell_1\cdot k_4,0,0,0,\ell_1\cdot k_4\right), 
\el
\left(0,0,0,\frac{s_{12}}{2},-\ell_1\cdot k_4,0,0\right), \left(0,0,0,-\frac{s_{12}}{2},0,0,0\right),\left(0,0,0,\frac{s_{12}}{2}+\ell_2\cdot k_1,-\frac{\chi_{14}s_{12}}{2}\right.
\el
\left. + \ell_2\cdot k_1,\ell_2 \cdot k_1, \ell_2\cdot k_1\right),\left(0,0,0,-\ell_2\cdot k_1,\frac{\chi_{14}s_{12}}{2}-\ell_2\cdot k_1,-\frac{s_{12}}{2}-\ell_2\cdot k_1,-\ell_2\cdot k_1\right),
\el
\left(0,0,0,-\frac{s_{12}}{2},0,0,\ell_1\cdot k_4\right)\bigg\}\,.
\label{gen1}
\eea
If we let $x_1 = \ell_1\cdot k_4$, $x_2 = \ell_2\cdot k_1$, $x_3 = s_{12}$, and take the dot product of each generator in (\ref{gen1}) with $\left(t_1,t_2,t_3,t_4,t_5,t_6,t_7\right)$, we arrive at the generating set for the ideal (hereafter referred to as $\mathcal{I}$) canonically associated to $\M$, $P_0$:
\bea
&&P_0 = \left\{-\frac{1}{2} t_3 x_3, \frac{t_3 x_3}{2}-t_2 x_2,t_7 x_2-\frac{t_3 x_3}{2}, \frac{t_1 x_3}{2}-t_7 x_2, -\frac{1}{2} t_2 x_3 \chi _{14}+t_1 x_1+t_2 x_1+t_3 x_1\right.
\el
\left.+t_7 x_1+\frac{t_3 x_3}{2}, \frac{t_4 x_3}{2}-t_5 x_1, -\frac{1}{2} t_4 x_3,-\frac{1}{2} t_5 x_3 \chi _{14}+t_4 x_2+t_5 x_2+t_6 x_2+t_7 x_2+\frac{t_4
   x_3}{2},\right.
\el
\left.\frac{1}{2} t_5 x_3 \chi _{14}-t_4 x_2-t_5 x_2-t_6 x_2-t_7 x_2-\frac{t_6 x_3}{2},t_7 x_1-\frac{t_4 x_3}{2}\right\}\,.
\label{polys}
\eea
Only the first ten rows of $E$ remain non-zero after the reduction is carried out. Provided that all $m$ propagator denominators are independent of one another (which is certainly true in our case), the $P_0$ we arrive at in this fashion will always be a minimal generating set for $\mathcal{I}$ which implies that, as assumed in Algorithm \ref{alg}, ${\rm SyzList}_0 = B_0 = \{{\bf 0}\}$. Actually, for most of steps of the algorithm the explicit form of $P_0$ is unimportant and we suppress it, writing instead $P_0 = \{p_1,p_2,p_3,p_4,p_5,p_6,p_7,p_8,p_9,p_{10}\}$.

We initialize $d$ to one and enter the While loop. Since $B_0 = \{{\bf 0}\}$, the algorithm directs us to the lower branch of the If statement. $M_1^{(3)} = \{x_1,x_2,x_3\}$, $P_0 = \{p_1,p_2,p_3,p_4,p_5,p_6,p_7,$
$p_8,p_9,p_{10}\}$, and
\cmb{-.7 cm}{0 cm}
\bea
&&P_1 = \left\{p_1 x_1,p_1 x_2,p_1 x_3,p_2 x_1, p_2 x_2, p_2 x_3, p_3 x_1, p_3 x_2, p_3 x_3, p_4 x_1, p_4 x_2, p_4 x_3, p_5 x_1, p_5 x_2, p_5 x_3,\right.
\el
\left.p_6 x_1,p_6 x_2,p_6 x_3, p_7 x_1, p_7 x_2, p_7 x_3, p_8 x_1, p_8 x_2, p_8 x_3, p_9 x_1, p_9 x_2, p_9 x_3, p_{10} x_1, p_{10} x_2, p_{10} x_3\right\}\nn
\eea
\cme
has ${10 (1+3-1)!\over 1!(3-1)!}= 30$ elements. To find $D$ (Subroutine \ref{sub2}) we have to introduce a sequence of 30 dummy variables
\bea
Y &=& \left\{y_1, y_2, y_3, y_4, y_5, y_6, y_7, y_8, y_9, y_{10}, y_{11}, y_{12}, y_{13}, y_{14}, y_{15}, y_{16},\right.
\el
\left.y_{17}, y_{18}, y_{19}, y_{20}, y_{21}, y_{22}, y_{23}, y_{24}, y_{25}, y_{26}, y_{27}, y_{28}, y_{29}, y_{30}\right\}\,,
\eea
expand the dot product $G \cdot Y$, and partition the terms in the resulting sum into equivalence classes. Two terms are equivalent if and only if they contain the same monomial in the variables $\{t_1,t_2,t_3,t_4,t_5,t_6,t_7,x_1,x_2,x_3\}$. Using the explicit form of the $p_i$ given in eq. (\ref{polys}), we find 37 equivalence classes:
\cmb{-.8 cm}{0 cm}
\bea
&&\left\{t_1 x_1^2 y_{13},t_2 x_1^2 y_{13},t_3 x_1^2 y_{13},-t_5 x_1^2 y_{16},t_7 y_{13} x_1^2+t_7 y_{28} x_1^2,t_1 x_1 x_2 y_{14},t_2 x_1 x_2 y_{14}-t_2 x_1 x_2 y_4,\right.
\el
\left.t_3 x_1 x_2 y_{14},t_4 x_1 x_2 y_{22}-t_4 x_1 x_2 y_{25},-t_5 x_1 x_2 y_{17}+t_5 x_1 x_2 y_{22}-t_5 x_1 x_2 y_{25},t_6 x_1 x_2 y_{22}-t_6 x_1 x_2 y_{25},\right.
\el
\left. t_7 x_1 x_2 y_7-t_7 x_1 x_2 y_{10}+t_7 x_1 x_2 y_{14}+t_7 x_1 x_2 y_{22}-t_7 x_1 x_2 y_{25}+t_7 x_1 x_2 y_{29},-t_2 x_2^2 y_5,t_4 x_2^2 y_{23}\right.
\el
\left.-t_4 x_2^2 y_{26},t_5 x_2^2 y_{23}-t_5 x_2^2 y_{26},t_6 x_2^2 y_{23}-t_6 x_2^2 y_{26},t_7 y_8 x_2^2-t_7 y_{11} x_2^2+t_7 y_{23} x_2^2-t_7 y_{26} x_2^2,\right.
\el
\left.\frac{1}{2} t_1 x_1 x_3
   y_{10}+t_1 x_1 x_3 y_{15},t_2 x_1 x_3 y_{15}-\frac{1}{2} t_2 x_1 x_3 y_{13} \chi _{14},-\frac{1}{2} t_3 x_1 x_3 y_1+\frac{1}{2} t_3 x_1 x_3
   y_4\right.
\el
\left.-\frac{1}{2} t_3 x_1 x_3 y_7+\frac{1}{2} t_3 x_1 x_3 y_{13}+t_3 x_1 x_3 y_{15},\frac{1}{2} t_4 x_1 x_3 y_{16}-\frac{1}{2} t_4 x_1 x_3
   y_{19}+\frac{1}{2} t_4 x_1 x_3 y_{22}\right.
\el
\left.-\frac{1}{2} t_4 x_1 x_3 y_{28},-t_5 x_1 x_3 y_{18}-\frac{1}{2} t_5 x_1 x_3 y_{22} \chi _{14}+\frac{1}{2} t_5
   x_1 x_3 y_{25} \chi _{14},-\frac{1}{2} t_6 x_1 x_3 y_{25},t_7 x_1 x_3 y_{15}\right.
\el
\left.+t_7 x_1 x_3 y_{30},\frac{1}{2} t_1 x_2 x_3 y_{11},-t_2 x_2 x_3
   y_6-\frac{1}{2} t_2 x_2 x_3 y_{14} \chi _{14},-\frac{1}{2} t_3 x_2 x_3 y_2+\frac{1}{2} t_3 x_2 x_3 y_5\right.
\el
\left.-\frac{1}{2} t_3 x_2 x_3 y_8+\frac{1}{2} t_3
   x_2 x_3 y_{14},\frac{1}{2} t_4 x_2 x_3 y_{17}-\frac{1}{2} t_4 x_2 x_3 y_{20}+\frac{1}{2} t_4 x_2 x_3 y_{23}+t_4 x_2 x_3 y_{24}\right.
\el
\left.-t_4 x_2 x_3
   y_{27}-\frac{1}{2} t_4 x_2 x_3 y_{29},t_5 x_2 x_3 y_{24}-t_5 x_2 x_3 y_{27}-\frac{1}{2} t_5 x_2 x_3 y_{23} \chi _{14}+\frac{1}{2} t_5 x_2 x_3
   y_{26} \chi _{14},\right.
\el
\left.t_6 x_2 x_3 y_{24}-\frac{1}{2} t_6 x_2 x_3 y_{26}-t_6 x_2 x_3 y_{27},t_7 x_2 x_3 y_9-t_7 x_2 x_3 y_{12}+t_7 x_2 x_3
   y_{24}-t_7 x_2 x_3 y_{27},\right.
\el
\left.\frac{1}{2} t_1 x_3^2 y_{12},-\frac{1}{2} t_2 x_3^2 y_{15} \chi _{14},-\frac{1}{2} t_3 y_3 x_3^2+\frac{1}{2} t_3 y_6 x_3^2-\frac{1}{2} t_3 y_9 x_3^2+\frac{1}{2} t_3 y_{15} x_3^2,\frac{1}{2} t_4 y_{18} x_3^2 \right.
\el
\left.-\frac{1}{2} t_4 y_{21} x_3^2+\frac{1}{2} t_4 y_{24}
   x_3^2-\frac{1}{2} t_4 y_{30} x_3^2,\frac{1}{2} t_5 x_3^2 y_{27} \chi _{14}-\frac{1}{2} t_5 x_3^2 y_{24} \chi _{14},-\frac{1}{2} t_6 x_3^2 y_{27}\right\}\,.
\eea
\cme
After setting the terms in each equivalence class to zero, we find that six of the resulting equations are obviously redundant. The remaining 31 equations,
\bea
&&\left\{0=y_5,0=y_{11},0=y_{12},0=y_{13},0=y_4-y_{14},0=y_2-y_5+y_8-y_{14},0=y_{14},\right.
\el
\left.0=\frac{y_{14} \chi_{14}}{2}+y_6,0=y_1-y_4+y_7-y_{13}-2
   y_{15},0=y_3-y_6+y_9-y_{15},0=y_{13}-\frac{2 y_{15}}{\chi _{14}},\right.
\el
\left.0=y_{15},0=y_{10}+2 y_{15},0=y_{16},0=\frac{y_{22} \chi _{14}}{2}-\frac{y_{25} \chi_{14}}{2}+y_{18},0=y_{22}-y_{25},0=y_{25},\right.
\el
\left.0=y_{17}-y_{22}+y_{25},0=y_{23}-y_{26},0=y_8-y_{11}+y_{23}-y_{26},0=y_{24}-y_{27},\right.
\el
\left.0=y_9-y_{12}+y_{24}-y_{27},0=y_{24}-\frac{y_{26}}{2}-y_{27},0=-\frac{2 y_{24}}{\chi _{14}}+\frac{2 y_{27}}{\chi_{14}}+y_{23}-y_{26},0=y_{27},\right.
\el
\left.0=y_{16}-y_{19}+y_{22}-y_{28},0=y_{13}+y_{28},0=y_{17}-y_{20}+y_{23}+2 y_{24}-2
   y_{27}-y_{29},\right.
\el
\left.0=y_7-y_{10}+y_{14}+y_{22}-y_{25}+y_{29},0=y_{18}-y_{21}+y_{24}-y_{30},0=y_{15}+y_{30}\right\}\,,
\label{eqs1}
\eea
constrain the $y_i$. After putting (\ref{eqs1}) into matrix form, we can easily solve for the null space of the resulting matrix. We find that, as expected, the null space is one-dimensional:
\be
D = \Big\{\left(1,0,0,0,0,0,-1,0,0,0,0,0,0,0,0,0,0,0,0,-1,0,0,0,0,0,0,0,0,1,0\right)\Big\}.
\ee
We have now determined a basis for the scalar syzygies of $P_1$:
\be
B_1 = D
\ee
To determine what syzygy of $P_0$ this element of $S_0(P_1)$ corresponds to, we apply the same mapping that we used in Section \ref{review} (formalized in Subroutine \ref{sub3}) in the example illustrating why computing a basis for a syzygy module with linear algebra is non-trivial. We partition the single element of $D$ into ten distinct non-overlapping sequences of length three without disturbing the ordering of the vector that we started with
\be
\Big\{\{1,0,0\},\{0,0,0\},\{-1,0,0\},\{0,0,0\},\{0,0,0\},\{0,0,0\},\{0,-1,0\},\{0,0,0\},\{0,0,0\},\{0,1,0\}\Big\}
\ee
and then take the dot product of each sequence of three elements with $M_1^{(3)} = \{x_1,x_2,x_3\}$:
\be
{\rm SyzList}_1 = \bigg\{\Big(x_1,0,-x_1,0,0,0,-x_2,0,0,x_2\Big)\bigg\}\,.
\ee
It is easy to check that $(x_1,0,-x_1,0,0,0,-x_2,0,0,x_2)$ is indeed a syzygy of $P_0$.

Now we increment $d$ to two and begin our second pass through the While loop. This time, $B_1 \neq \{{\bf 0}\}$ and, therefore, the algorithm directs us to the upper branch of the If statement.  $M_1^{(3)} = \{x_1,x_2,x_3\}$, $P_0 = \{p_1,p_2,p_3,p_4,p_5,p_6,p_7,p_8,p_9,p_{10}\}$, and
\cmb{-.7 cm}{0 cm}
\bea
&&P_1 = \left\{p_1 x_1,p_1 x_2,p_1 x_3,p_2 x_1, p_2 x_2, p_2 x_3, p_3 x_1, p_3 x_2, p_3 x_3, p_4 x_1, p_4 x_2, p_4 x_3, p_5 x_1, p_5 x_2, p_5 x_3,\right.
\el
\left.p_6 x_1,p_6 x_2,p_6 x_3, p_7 x_1, p_7 x_2, p_7 x_3, p_8 x_1, p_8 x_2, p_8 x_3, p_9 x_1, p_9 x_2, p_9 x_3, p_{10} x_1, p_{10} x_2, p_{10} x_3\right\}\nn
\eea
\cme
with ${10 (1+3-1)!\over 1!(3-1)!}= 30$ elements. To determine $P_2$, we need the set of monomials of degree two built out of $\{x_1,x_2,x_3\}$, $M_2^{(3)} = \left\{x_1^2,x_1 x_2, x_1 x_3, x_2^2, x_2 x_3, x_3^2\right\}$. As mentioned in Section \ref{body}, for definiteness, we have chosen the lexicographical ordering for our sets of monomials. Let us say again that this choice is not by any means necessary. A different choice of ordering for the monomials would lead to a different representation of the same syzygy module; the number of linearly independent syzygies at each degree would be exactly the same. Continuing, we see that
\bea
&&P_2 = \left\{p_1 x_1^2,p_1 x_1 x_2,p_1 x_1 x_3,p_1 x_2^2,p_1 x_2 x_3,p_1 x_3^2,p_2 x_1^2,p_2 x_1 x_2,p_2 x_1 x_3,p_2 x_2^2,p_2 x_2 x_3,p_2 x_3^2,\right.
\el
\left.p_3 x_1^2,p_3 x_1 x_2,p_3 x_1 x_3,p_3 x_2^2,p_3 x_2 x_3,p_3 x_3^2,p_4 x_1^2,p_4 x_1 x_2,p_4 x_1 x_3,p_4 x_2^2,p_4 x_2 x_3,p_4 x_3^2,\right.
\el
\left.p_5 x_1^2,p_5 x_1 x_2,p_5 x_1 x_3,p_5 x_2^2,p_5 x_2 x_3,p_5 x_3^2,p_6 x_1^2,p_6 x_1 x_2,p_6 x_1 x_3,p_6 x_2^2,p_6 x_2 x_3,p_6 x_3^2,\right.
\\
&&\left.p_7 x_1^2,p_7 x_1 x_2,p_7 x_1 x_3,p_7 x_2^2,p_7 x_2 x_3,p_7 x_3^2,p_8 x_1^2,p_8 x_1 x_2,p_8 x_1 x_3,p_8 x_2^2,p_8 x_2 x_3,p_8 x_3^2,\right.
\el
\left.p_9 x_1^2,p_9 x_1 x_2,p_9 x_1 x_3,p_9 x_2^2,p_9 x_2 x_3,p_9 x_3^2,p_{10} x_1^2,p_{10} x_1 x_2,p_{10} x_1 x_3,p_{10} x_2^2,p_{10} x_2 x_3,p_{10} x_3^2\right\}\nonumber
\eea
has ${10 (2+3-1)!\over 2!(3-1)!}= 60$ elements. $P_1$ is the basis with respect to which the single element of $B_1$ has components
\be
\left(1,0,0,0,0,0,-1,0,0,0,0,0,0,0,0,0,0,0,0,-1,0,0,0,0,0,0,0,0,1,0\right)\,.
\label{deg1syz}
\ee
The function $\sigma$ (Subroutine \ref{sub1}) maps this vector to three new vectors in the larger vector space for which $P_2$
is the standard basis. This map is carried out by extending the $P_1$ by each of the three variables in turn. Applying $\sigma$ to 
\bea
&&\left(1,0,0,0,0,0,-1,0,0,0,0,0,0,0,0,0,0,0,0,-1,0,0,0,0,0,0,0,0,1,0\right)
\el
\qquad\qquad\qquad= p_1 x_1-p_3 x_1-p_7 x_2+p_{10} x_2\,,
\eea
we obtain
\bea
&&\sigma\left(p_1 x_1-p_3 x_1-p_7 x_2+p_{10} x_2\right) = \left\{p_1 x_1^2-p_3 x_1^2-p_7 x_1 x_2+p_{10}x_1 x_2,\right.
\el
\left.p_1 x_1 x_2-p_3 x_1 x_2-p_7 x_2^2+p_{10}x_2^2, p_1 x_1 x_3-p_3 x_1 x_3-p_7 x_2 x_3+p_{10}x_2 x_3\right\}\,.
\eea
We can now read off the components of three scalar syzygies of $P_2$ which have their origin in the scalar syzygy of $P_1$ that we computed on our first pass through the While loop. They are:
\cmb{-.7 cm}{0 cm}
\bea
&&A =\Big\{(1,0,0,0,0,0,0,0,0,0,0,0,-1,0,0,0,0,0,0,0,0,0,0,0,0,0,0,0,0,0,0,0,0,0,0,0,0,
\el
-1,0,0,0,0,0,0,0,0,0,0,0,0,0,0,0,0,0,1,0,0,0,0), (0,1,0,0,0,0,0,0,0,0,0,0,0,-1,0,0,
\el
0,0,0,0,0,0,0,0,0,0,0,0,0,0,0,0,0,0,0,0,0,0,0,-1,0,0,0,0,0,0,0,0,0,0,0,0,0,0,0,0,
\el
0,1,0,0), (0,0,1,0,0,0,0,0,0,0,0,0,0,0,-1,0,0,0,0,0,0,0,0,0,0,0,0,0,0,0,0,0,0,0,0,0,
\el
0,0,0,0,-1,0,0,0,0,0,0,0,0,0,0,0,0,0,0,0,0,0,1,0)\Big\}\,.
\eea
\cme
If we view the elements of $A$ as the rows of a matrix, the result is already in row echelon form. It follows that $C$ is simply $A$ in matrix form. The first three columns of $C$ are pivot columns, so $G$ is $P_2$ with the first three entries replaced with zero:
\cmb{-.6 cm}{0 cm}
\bea
&&G = \left\{0,0,0,p_1 x_2^2,p_1 x_2 x_3,p_1 x_3^2,p_2 x_1^2,p_2 x_1 x_2,p_2 x_1 x_3,p_2 x_2^2,p_2 x_2 x_3,p_2 x_3^2,p_3 x_1^2,\right.
\el
\left.p_3 x_1 x_2,p_3 x_1 x_3,p_3 x_2^2,p_3 x_2 x_3,p_3 x_3^2,p_4 x_1^2,p_4 x_1 x_2,p_4 x_1 x_3,p_4 x_2^2,p_4 x_2 x_3,p_4 x_3^2,p_5 x_1^2,\right.
\el
\left.p_5 x_1 x_2,p_5 x_1 x_3,p_5 x_2^2,p_5 x_2 x_3,p_5 x_3^2,p_6 x_1^2,p_6 x_1 x_2,p_6 x_1 x_3,p_6 x_2^2,p_6 x_2 x_3,p_6 x_3^2,p_7 x_1^2,\right.
\\
&&\left.p_7 x_1 x_2,p_7 x_1 x_3,p_7 x_2^2,p_7 x_2 x_3,p_7 x_3^2,p_8 x_1^2,p_8 x_1 x_2,p_8 x_1 x_3,p_8 x_2^2,p_8 x_2 x_3,p_8 x_3^2,p_9 x_1^2,\right.
\el
\left.p_9 x_1 x_2,p_9 x_1 x_3,p_9 x_2^2,p_9 x_2 x_3,p_9 x_3^2,p_{10} x_1^2,p_{10} x_1 x_2,p_{10} x_1 x_3,p_{10} x_2^2,p_{10} x_2 x_3,p_{10} x_3^2\right\}\nonumber\,.
\eea
\cme
Making this replacement will prevent Subroutine \ref{sub2} from rediscovering syzygies that have their origin in the syzygy of $P_1$ that we found on our first pass through the While loop. On this pass through the loop, we refrain from stepping into Subroutines \ref{sub2} and \ref{sub3} since their functionality should already be quite clear from our first pass through. Applying ${\rm SSyz}$ to $G$, we find
\cmb{-.5 cm}{0 cm}
\bea
&&D = {\rm SSyz}(G) = \bigg\{\Big(0,0,0,0,0,0,0,0,0,0,0,0,0,0,0,0,0,0,0,0,0,0,0,0,0,0,0,0,0,0,0,0,
\el
0,0,-1,\frac{\chi_{14}}{2},0,-4,-1,0,-2,\frac{\chi_{14}}{2},0,-2,-1,0,0,0,0,-2,0,0,0,0,0,0,0,0,1,0\Big), \Big(0,0,0,0,
\el
-\frac{1}{4},\frac{\chi_{14}}{8},0,0,-\frac{1}{4},0,0,\frac{\chi_{14}}{8},0,-\frac{1}{2},-\frac{1}{4},0,0,0,0,\frac{1}{2},0,0,0,0,0,0,0,0,-\frac{1}{4},0,0,0,0,0,\frac{1}{2},-\frac{\chi_{14}}{4},
\el
0,2,\frac{1}{2},-1,\frac{1}{2},-\frac{\chi _{14}}{4},0,1,\frac{1}{2},0,0,0,0,1,0,0,0,0,0,0,0,1,0,0\Big)\bigg\}
\eea
\cme
for the new scalar syzygies of $P_2$. Since $D \neq \{{\bf 0}\}$,
\cmb{-.5 cm}{0 cm}
\bea
&&B_2 = C \cup D = \Big\{(1,0,0,0,0,0,0,0,0,0,0,0,-1,0,0,0,0,0,0,0,0,0,0,0,0,0,0,0,0,0,0,0,0,
\el
0,0,0,0,-1,0,0,0,0,0,0,0,0,0,0,0,0,0,0,0,0,0,1,0,0,0,0), (0,1,0,0,0,0,0,0,0,0,0,0,0,
\el
-1,0,0,0,0,0,0,0,0,0,0,0,0,0,0,0,0,0,0,0,0,0,0,0,0,0,-1,0,0,0,0,0,0,0,0,0,0,0,0,0,
\el
0,0,0,0,1,0,0), (0,0,1,0,0,0,0,0,0,0,0,0,0,0,-1,0,0,0,0,0,0,0,0,0,0,0,0,0,0,0,0,0,0,
\el
0,0,0,0,0,0,0,-1,0,0,0,0,0,0,0,0,0,0,0,0,0,0,0,0,0,1,0), \Big(0,0,0,0,0,0,0,0,0,0,0,0,0,
\el
0,0,0,0,0,0,0,0,0,0,0,0,0,0,0,0,0,0,0,0,0,-1,\frac{\chi_{14}}{2},0,-4,-1,0,-2,\frac{\chi_{14}}{2},0,-2,-1,0,0,0,
\el
0,-2,0,0,0,0,0,0,0,0,1,0\Big),\Big(0,0,0,0,-\frac{1}{4},\frac{\chi_{14}}{8},0,0,-\frac{1}{4},0,0,\frac{\chi_{14}}{8},0,-\frac{1}{2},-\frac{1}{4},0,0,0,0,\frac{1}{2},0,
\el
0,0,0,0,0,0,0,-\frac{1}{4},0,0,0,0,0,\frac{1}{2},-\frac{\chi_{14}}{4},0,2,\frac{1}{2},-1,\frac{1}{2},-\frac{\chi_{14}}{4},0,1,\frac{1}{2},0,0,0,0,1,0,0,0,0,0,0,
\el
0,1,0,0\Big)\Big\}\,.
\eea
\cme
The two new scalar syzygies of $P_2$ discovered by {\rm SSyz} map, via {\rm ToSyz}, to two new syzygies of $P_0$ of degree two, linearly independent of $(x_1,0,-x_1,0,0,0,-x_2,0,0,x_2)$:
\bea
&&{\rm SyzList}_2 = {\rm ToSyz}(D) = \bigg\{\Big(0,0,0,0,0,\frac{1}{2} x_3^2 \chi _{14}-x_2 x_3,\frac{1}{2} x_3^2 \chi _{14}-x_1 x_3-2 x_2 x_3
\el
-4 x_1 x_2,-2 x_1 x_2-x_1 x_3,-2 x_1 x_2,x_2 x_3\Big),\Big(\frac{1}{8} x_3^2 \chi _{14}-\frac{x_2 x_3}{4},\frac{1}{8} x_3^2 \chi _{14}-\frac{x_1 x_3}{4},
\el
-\frac{1}{2} x_1 x_2-\frac{x_1 x_3}{4},\frac{x_1 x_2}{2},-\frac{1}{4} x_2 x_3,\frac{x_2 x_3}{2}-\frac{1}{4} x_3^2 \chi _{14},-\frac{1}{4} x_3^2 \chi_{14}-x_2^2+2 x_1 x_2
\el
+\frac{x_3 x_2}{2}+\frac{x_1 x_3}{2},x_1 x_2+\frac{x_1 x_3}{2},x_1 x_2,x_2^2\Big)\bigg\}\,.
\eea
It is again simple to check that the elements of ${\rm SyzList}_2$ are in fact syzygies of $P_0$. If we did not have {\it a priori} knowledge of the syzygy module, we would have to continue on and pass through the While loop several more times (as explained in~\cite{GKK}, $\Delta = 6$ for the planar massless double box) to search for more potentially useful linearly independent syzygies.\footnote{It is important to realize that, in general, $D = \{{\bf 0}\}$ for $d = i$ does {\it not} imply that $D = \{{\bf 0}\}$ for $d = i + 1$, though this may turn out to be true in practice for physically motivated examples.} Since we do not expect to find further solutions, we can exit the While loop after just two iterations and collect the results:
\bea
&&\bigcup\limits_{d=1}^2 {\rm SyzList}_d = \bigg\{\Big(x_1,0,-x_1,0,0,0,-x_2,0,0,x_2\Big),\Big(0,0,0,0,0,\frac{1}{2} x_3^2 \chi _{14}-x_2 x_3,\frac{1}{2} x_3^2 \chi_{14}
\el
-x_1 x_3-2 x_2 x_3-4 x_1 x_2,-2 x_1 x_2-x_1 x_3,-2 x_1 x_2,x_2 x_3\Big),\Big(\frac{1}{8} x_3^2 \chi _{14}-\frac{x_2 x_3}{4},\frac{1}{8} x_3^2 \chi _{14}
\el
-\frac{x_1 x_3}{4},-\frac{1}{2} x_1 x_2-\frac{x_1 x_3}{4},\frac{x_1 x_2}{2},-\frac{1}{4} x_2 x_3,\frac{x_2 x_3}{2}-\frac{1}{4} x_3^2 \chi _{14},-\frac{1}{4} x_3^2 \chi_{14}-x_2^2+2 x_1 x_2
\el
+\frac{x_3x_2}{2}+\frac{x_1 x_3}{2},x_1 x_2+\frac{x_1 x_3}{2},x_1 x_2,x_2^2\Big)\bigg\}\,.
\eea
Before leaving this section, we translate the above result back into the usual language used to describe the planar massless double box:
\cmb{-.7 cm}{0 cm}
\bea
&&\bigcup\limits_{d=1}^2 {\rm SyzList}_d = \bigg\{\Big(\ell_1 \cdot k_4,0,-\ell_1 \cdot k_4,0,0,0,-\ell_2 \cdot k_1,0,0,\ell_2 \cdot k_1\Big),\Big(0,0,0,0,0,\frac{1}{2} s_{12}^2 \chi _{14}
\\
&&-\ell_2\cdot k_1\,s_{12},\frac{1}{2} s_{12}^2 \chi_{14}-\ell_1 \cdot k_4 \,s_{12}-2 \ell_2 \cdot k_1\, s_{12}-4 \ell_1 \cdot k_4\, \ell_2 \cdot k_1,-2 \ell_1 \cdot k_4\, \ell_2 \cdot k_1-\ell_1 \cdot k_4\, s_{12},
\el
-2 \ell_1 \cdot k_4\, \ell_2 \cdot k_1,\ell_2 \cdot k_1\, s_{12}\Big),\Big(\frac{1}{8} s_{12}^2 \chi _{14}-\frac{\ell_2 \cdot k_1\, s_{12}}{4},\frac{1}{8} s_{12}^2 \chi _{14}-\frac{\ell_1 \cdot k_4\, s_{12}}{4},-\frac{1}{2} \ell_1 \cdot k_4 \,\ell_2 \cdot k_1
\el
-\frac{\ell_1 \cdot k_4\, s_{12}}{4},\frac{\ell_1 \cdot k_4 \,\ell_2 \cdot k_1}{2},-\frac{1}{4} \ell_2 \cdot k_1\, s_{12},\frac{\ell_2 \cdot k_1\, s_{12}}{2}-\frac{1}{4} s_{12}^2 \chi _{14},-\frac{1}{4} s_{12}^2 \chi_{14}-\left(\ell_2 \cdot k_1\right)^2
\nn
&&+2 \ell_1 \cdot k_4\, \ell_2 \cdot k_1+\frac{s_{12}\,\ell_2 \cdot k_1}{2}+\frac{\ell_1 \cdot k_4\, s_{12}}{2},\ell_1 \cdot k_4\, \ell_2 \cdot k_1+\frac{\ell_1 \cdot k_4\, s_{12}}{2},\ell_1 \cdot k_4\,\ell_2 \cdot k_1,\left(\ell_2 \cdot k_1\right)^2\Big)\bigg\}\,. \nonumber
\eea
\cme
\section{Conclusions}
\label{ConclusionSection}
This paper continued the program of research initiated in~\cite{GKK} by Gluza, Kajda, and Kosower (GKK). In reference~\cite{GKK}, GKK proposed a novel reduction scheme for multi-loop integrals guaranteed to produce unitarity-compatible integral bases free of doubled propagator denominators. In this work we took a fresh look at the computationally intensive part of their procedure, the generation of complete sets of unitarity-compatible integration by parts relations. Drawing upon some of the ideas in reference~\cite{LASyz}, we found an attractive alternative to the GKK approach which completely avoids the use of Gr\"{o}bner bases. In fact, we showed in Section \ref{body} that our solution, Algorithm \ref{alg}, can be described in terms of simple linear algebra. 

One shortcoming of the present paper is that we cannot yet claim to have fully optimized Algorithm \ref{alg}. Even if we excise the trivial redundancies introduced into our pseudo-code for the sake of clarity, there are many features of the algorithm which might benefit from further optimization. For example, as should already be clear from the non-trivial example discussed in Section \ref{example}, the matrices produced by Algorithm \ref{alg} are typically quite sparse and, so far, we have made no attempt to exploit this feature of the problem. Furthermore, it seems likely that a high-level implementation in {\tt Mathematica} (such as the one written by the author) will ultimately prove insufficient for research purposes. Besides the fact that, as a general rule, {\tt Mathematica} runs very slowly, it is not at all clear that {\tt Mathematica} exploits the best available algorithms for row reducing matrices; if our preliminary experimentations are any guide, it appears that {\tt Mathematica} manages system resources rather poorly. Fortunately, we anticipate that the brevity and simplicity of Algorithm \ref{alg} will make it possible to optimize and then effectively implement at a lower level in {\tt C++} or {\tt Fortran}. We are excited by this prospect and hope to pursue a project along these lines in the near future. 

\section*{Acknowledgments}

I would like to thank Janusz Gluza and David Kosower for inspiring me to pursue this line of research. I am also very grateful to Valery Yundin for his critical reading of an earlier draft of this work. The {\tt LaTeX} packages {\sc algorithm} and {\sc algorithmic}, written by Peter Williams and maintained by Roger\'{i}o Brito, were used to typeset Algorithm \ref{alg} and Subroutines \ref{sub1}, \ref{sub2}, and \ref{sub3}. Finally, I gratefully acknowledge CICYT support through the project FPA-2009-09017 and CAM support through the project HEPHACOS S2009/ESP-1473.

\bibliographystyle{JHEP}
\bibliography{intred}

\providecommand{\href}[2]{#2}\begingroup\raggedright\begin{thebibliography}{10}

\bibitem{Chetyrkin}
K.~G. Chetyrkin and F.~V. Tkachov, {\it {Integration by Parts: The Algorithm to
  Calculate beta Functions in 4 Loops}},  {\em Nucl. Phys.} {\bf B192} (1981)
  159--204.

\bibitem{EFInts}
V.~A. Smirnov, {\it {Evaluating Feynman integrals}},  {\em Springer Tracts Mod.
  Phys.} {\bf 211} (2004) 1--244.

\bibitem{MINCER}
S.~G. Gorishnii, S.~A. Larin, L.~R. Surguladze, and F.~V. Tkachov, {\it
  {MINCER: PROGRAM FOR MULTILOOP CALCULATIONS IN QUANTUM FIELD THEORY FOR THE
  SCHOONSCHIP SYSTEM}},  {\em Comput. Phys. Commun.} {\bf 55} (1989) 381--408.

\bibitem{Laporta}
S.~Laporta, {\it {High-precision calculation of multi-loop Feynman integrals by
  difference equations}},  {\em Int. J. Mod. Phys.} {\bf A15} (2000)
  5087--5159, [\href{http://xxx.lanl.gov/abs/hep-ph/0102033}{{\tt
  hep-ph/0102033}}].

\bibitem{AIR}
C.~Anastasiou and A.~Lazopoulos, {\it {Automatic integral reduction for higher
  order perturbative calculations}},  {\em JHEP} {\bf 07} (2004) 046,
  [\href{http://xxx.lanl.gov/abs/hep-ph/0404258}{{\tt hep-ph/0404258}}].

\bibitem{FIRE}
A.~V. Smirnov, {\it {Algorithm FIRE -- Feynman Integral REduction}},  {\em
  JHEP} {\bf 10} (2008) 107, [\href{http://xxx.lanl.gov/abs/0807.3243}{{\tt
  arXiv:0807.3243}}].

\bibitem{REDUZE}
C.~Studerus, {\it {Reduze - Feynman Integral Reduction in C++}},  {\em Comput.
  Phys. Commun.} {\bf 181} (2010) 1293--1300,
  [\href{http://xxx.lanl.gov/abs/0912.2546}{{\tt arXiv:0912.2546}}].

\bibitem{KLunitarity}
D.~A. Kosower and K.~J. Larsen, {\it {Maximal Unitarity at Two Loops}},
  \href{http://xxx.lanl.gov/abs/1108.1180}{{\tt arXiv:1108.1180}}.

\bibitem{MOunitarity}
P.~Mastrolia and G.~Ossola, {\it {On the Integrand-Reduction Method for
  Two-Loop Scattering Amplitudes}},
  \href{http://xxx.lanl.gov/abs/1107.6041}{{\tt arXiv:1107.6041}}.

\bibitem{GKK}
J.~Gluza, K.~Kajda, and D.~A. Kosower, {\it {Towards a Basis for Planar
  Two-Loop Integrals}},  {\em Phys. Rev.} {\bf D83} (2011) 045012,
  [\href{http://xxx.lanl.gov/abs/1009.0472}{{\tt arXiv:1009.0472}}].

\bibitem{alggeom}
D.~{Bayer} and D.~{Mumford}, {\it {What can be computed in algebraic
  geometry?}},  \href{http://xxx.lanl.gov/abs/alg-geom/9304003}{{\tt
  alg-geom/9304003}}.

\bibitem{AL}
W.~Adams and P.~Loustaunau, {\em {An Introduction to Gr\"{o}bner Bases}}.
\newblock American Mathematical Society, Graduate Studies in Mathematics,
  Volume 3, 1994.

\bibitem{LASyz}
D.~{Cabarcas} and J.~{Ding}, {\it {Linear Algebra to Compute Syzygies and
  Gr\"{o}bner Bases}},  in {\em ISSAC '11: Proceedings of the 36th
  International Symposium on Symbolic and Algebraic Computation}, 2011.

\bibitem{Buchberger}
B.~{Buchberger}, ``\it {E}in {A}lgorithmus zum {A}uffinden der {B}asiselemente
  des {R}estklassenringes nach einem nulldimensionalen {P}olynomideal.'' \rm
  http://www.risc.jku.at/Groebner-Bases-Bibliography/details.php?details$\_$\,id=706.
\newblock \rm Ph. {D}\, dissertation, Universit\"at Innsbruck, 1965.

\bibitem{Eder}
C.~{Eder}, J.~{Gash}, and J.~{Perry}, {\it {Modifying Faug\`{e}re's F5
  Algorithm to ensure termination}},
  \href{http://xxx.lanl.gov/abs/1006.0318}{{\tt arXiv:1006.0318}}.

\bibitem{Faugere}
J.~{C.~Faug\`{e}re}, {\it {A New Efficient Algorithm For Computing Gr\"{o}bner
  Bases Without Reduction to Zero $(F_5)$}},  in {\em ISSAC '02: Proceedings of
  the 2002 International Symposium on Symbolic and Algebraic Computation},
  2002.

\bibitem{SyzF5}
G.~Ars and A.~Hashemi, {\it {C}omputing {S}yzygies by {F}aug\`{e}re's ${F}_5$
  {A}lgorithm},  {\em Results in Mathematics} {\bf 59} (2011) 35--42.

\bibitem{Stanley}
R.~Stanley, {\em {Enumerative Combinatorics, Volume I}}.
\newblock Cambridge University Press, 1997.

\end{thebibliography}\endgroup
\end{document}